\documentclass{Interspeech2024}

\usepackage[utf8]{inputenc}
\usepackage{CJK}
\usepackage{hyperref}

\usepackage{booktabs}
\usepackage{multirow}
\usepackage{footmisc}
\usepackage[table,xcdraw]{xcolor}

\interspeechcameraready 

\title{AS-70: A Mandarin stuttered speech dataset for automatic speech recognition and stuttering event detection}

\name[affiliation={1*}]{Rong}{Gong}
\name[affiliation={2*}]{Hongfei}{Xue}
\name[affiliation={1}]{Lezhi}{Wang}
\name[affiliation={3}]{Xin}{Xu}
\name[affiliation={4}]{Qisheng}{Li}
\name[affiliation={2}]{Lei}{Xie}
\name[affiliation={3}]{Hui}{Bu}
\name[affiliation={4}]{Shaomei}{Wu}
\name[affiliation={5}]{Jiaming}{Zhou}
\name[affiliation={5}]{Yong}{Qin}
\name[affiliation={6}]{Binbin}{Zhang}
\name[affiliation={7}]{Jun}{Du}
\name[affiliation={1}]{Jia}{Bin}
\name[affiliation={8}]{Ming}{Li}

\address{
  $^1$StammerTalk.
  $^2$ASLP@NPU, Northwestern Polytechnical University.
  $^3$AIShell Inc \\
  $^4$AImpower.
  $^5$Nankai University.
  $^6$WeNet Open Source Community \\
  $^7$University of Science and Technology of China.
  $^8$Duke Kunshan University}
\email{rong.gong@stammertalk.net, hfxue@mail.nwpu.edu.cn\thanks{*Equal contribution.}}


\begin{document}

\maketitle

\begin{abstract}
The rapid advancements in speech technologies over the past two decades have led to human-level performance in tasks like automatic speech recognition (ASR) for fluent speech. However, the efficacy of these models diminishes when applied to atypical speech, such as stuttering. This paper introduces AS-70, the first publicly available Mandarin stuttered speech dataset, which stands out as the largest dataset in its category. Encompassing conversational and voice command reading speech, AS-70 includes verbatim manual transcription, rendering it suitable for various speech-related tasks. Furthermore, baseline systems are established, and experimental results are presented for ASR and stuttering event detection (SED) tasks. By incorporating this dataset into the model fine-tuning, significant improvements in the state-of-the-art ASR models, e.g., Whisper and Hubert, are observed, enhancing their inclusivity in addressing stuttered speech.
\end{abstract}

\section{Introduction}

Stuttering is a speech impediment that affects around 1\% of the world's population~\cite{yairi_1996}. The impact of the stuttering on people who have such a problem could be on both social functioning and mental aspects. Stuttering hinders daily oral communication with speech repetition, prolongation, blocks, and secondary behaviors, such as body movements and facial grimaces~\cite{stutteringoverview}.

Early children who stutter (CWS) have around 80\% of a chance to recover within four years of the onset~\cite{Yairi_1999}. Early intervention by speech-language therapists (SLP) is crucial for increasing the recovery rate~\cite{onslow_management}. Compared with North American or Western European countries where the practice of SLP has been evolving for quite some time, the related practices in Mainland China are still in their infancy. Due to the lack of professionals, many families of children who stutter cannot receive timely diagnosis. While adding more professionals, automatic stuttering diagnosis can also meet the needs of some families.

When a person's stuttering continues into adulthood, the chance of full recovery becomes very low, and there is a high probability that stuttering will accompany them throughout their lives~\cite{self-reported}. To alleviate their communication barriers or, in a broad sense, eliminate social discrimination, relevant products need to be designed to be inclusive to meet their needs.

With the rise of smart home devices and chatbot technology, e.g., Alexa and ChatGPT, voice-user interfaces have become indispensable tools in many of our lives. Although current ASR systems can handle fluent speech well~\cite{whisper}, they encounter difficulties in recognizing stuttered speech~\cite{Lea_chi_2023,Shonibare2022EnhancingAF}. The underlying reasons for this challenge may involve various factors, including insufficient data or a lack of awareness regarding the necessity to develop systems tailored for PWS.

\subsection{Related works}

Due to the highly sensitive and personal nature of the data, large-scale stuttering speech datasets collected by companies are typically not made openly accessible to the research community~\cite{Lea_chi_2023, Bob_euphonia_2021, analysis-and-tuning, Shonibare2022EnhancingAF, stutter-tts}. On the other hand, openly accessible datasets curated by academic researchers, such as FluencyBank~\cite{fluencybank} and UCLASS~\cite{uclass}, are often small and primarily intended for use in SLP resources. Larger open datasets, while available, may require additional completeness in terms of annotation and authenticity. For instance, the Sep-28k dataset~\cite{sep-28k}, comprising 28k 3-second podcast audio clips, only provides stuttering labels for the SED task but lacks accompanying text transcription. Meanwhile, the LibriStutter~\cite{libristutter} dataset includes text transcriptions, but it is artificially generated from recordings of fluent speech. Furthermore, it is noteworthy that all these open stuttering speech datasets are limited to two Western languages~\cite{bayerl_KSoFKasselState_2022}.

The primary focus of most ASR research on stuttered speech is directed towards predicting semantic rather than verbatim transcription. Lea et al.~\cite{Lea_chi_2023} observed enhanced performance by modifying the ASR decoder, specifically by increasing the language model weight and imposing a higher penalty for word insertions. Shonibare et al.~\cite{Shonibare2022EnhancingAF} introduced a frame-level stuttering classifier, termed Detect-and-Pass, designed for the RNNT ASR model. Zhang et al.~\cite{stutter-tts} demonstrated performance improvement through fine-tuning the ASR model using synthesized stuttering speech. Alharbi et al.~\cite{Alharbi2017AutomaticRO} utilized ASR to generate orthographic transcriptions that include stuttering words.

SED involves identifying instances of stuttering in speech audio clips. Research in this domain predominantly focused on four openly available datasets: Sep-28k~\cite{sep-28k}, FluencyBank~\cite{fluencybank}, LibriStutter~\cite{libristutter}, and KSoF~\cite{bayerl_KSoFKasselState_2022}. Different neural network architectures, including ConvLSTM~\cite{sep-28k}, Stutternet~\cite{stutternet}, and SE-Resnet~\cite{libristutter}, were developed specifically for this task. Additionally, machine learning techniques like multi-task learning~\cite{sep-28k,bayerl23_interspeech} were employed to enhance the accuracy of stuttering detection.

The subsequent sections of this paper showcase our primary contribution - the AS-70 dataset, which can be accessed from this download location\footnote{\url{https://www.aishelltech.com/AISHELL_6A}}. In Section~\ref{sec:dataset}, we elaborate on the data collection, annotation process, and conduct a descriptive analysis that compares AS-70 with other openly available stuttered speech datasets. In Section~\ref{sec:experiments}, we establish baseline systems for the ASR and SED tasks using the AS-70 dataset.

\vspace{-5pt}
\section{Dataset}\label{sec:dataset}

\subsection{Participants and recording sessions}

The AS-70 dataset was recorded between January 2023 and October 2023 by two of the authors, both adults who stutter (AWS) and native Mandarin speakers. A total of 70 native Mandarin AWS took part in the recording sessions, with 24 of them being female, establishing a male-to-female ratio of 1.9:1. There was no gender selection process as all willing candidates were accepted to participate in the recording sessions.


Each participant engaged in a recording session lasting up to one hour, comprising two parts: conversation and voice command reading. Conversations were conducted through online interviews using platforms like Zoom or Tencent Meet, aiming to capture spontaneous speech on diverse topics. The interviewer, one of the two authors, posed questions based on a prepared list, with the flexibility to introduce impromptu questions as needed.


In the voice command reading part, participants were tasked with reading a set of 200 commands, categorized into car navigation and smart home device interaction. To ensure variety, a new set of 200 commands was introduced for every 25 participants, resulting in a dataset featuring a total of 600 unique commands. Participants were encouraged to employ the Voluntary Stuttering technique, deliberately introducing stuttering. 


\subsection{Annotation}
The dataset annotation process involved 15 non-PWS annotators and 5 quality controllers (QCs). Among the 5 QCs, 4 are non-PWS, and 1 is the PWS author (QC0). All annotators and QCs are experienced Mandarin speech annotators, with some having prior experience annotating atypical speech. To ensure annotation quality, QC0 conducted a training session prior to the commencement of the annotation process. In this session, the 19 non-PWS annotators and QCs were initially tasked with annotating a few examples of stuttered speech. Following this, QC0 provided feedback, and the 19 annotators were then asked to annotate new examples, incorporating the lessons learned from the feedback. This iterative process was repeated multiple times until no further feedback was deemed necessary.

Five types of stuttering were specified by the annotation guidelines, including:
\begin{CJK*}{UTF8}{gbsn}
\begin{itemize}
\item \textbf{[]}: \textbf{Word/phrase repetition}. Designated for marking entire repeated character or phrase.
\item \textbf{/b}: \textbf{block}. Gasps for air or stuttered pauses.
\item \textbf{/p}: \textbf{prolongation}. Elongated phoneme.
\item \textbf{/r}: \textbf{sound repetition}. Repeated phoneme that do not constitute an entire character.
\item \textbf{/i}: \textbf{interjections}. Filler characters due to stuttering e.g., `嗯', `啊', or `呃'. Notably, naturally occurring interjections that don't disrupt the speech flow are excluded.
\end{itemize}

The annotation was performed in a verbatim manner, with stuttering labels embedded as markups. An example of the annotated transcription can be ``嗯/i/p，我[我我]的名/b字是小/r明。". The character `嗯' is an interjection and also is prolonged. The character `我' is repeated twice additionally. A block annotation is applied to the character `名'. Furthermore, sound repetition is identified in one of the phonemes of `小'. It is noteworthy that a single character may carry multiple labels. Additional examples of audio and annotations can be accessed through this link\footnote{\url{https://stammertalk.github.io/interspeech2024-page}\label{fn:webpage}}.
\end{CJK*}

The conversation and command reading parts of each session were annotated separately by different annotators. Subsequently, one of the 4 non-PWS QCs conducted a cross-check of the annotations. To ensure consistent annotation, QC0 then performed a final cross-check of all session annotations. The annotators noted that this annotation process required approximately three times more time compared to annotating fluent speech.

Personal Identifiable Information (PII) such as name, address, birth date, and workplace has been redacted from the transcription. The audio segments containing PII have been muted. A data collection agreement, adhering to Guidance
for personal information security impact assessment~\cite{GBT39335-2020}, was signed between the data collectors and the participant.

\subsection{Descriptive analysis}

\begin{table*}[ht!]
\caption{Dataset scale and scope as characterized by speech duration (Duration), the number and types of speakers (Speakers), whether it provides speech transcription (Transcription), types of speaking tasks (Tasks), and Language. * Limited to the transcribed portion of the dataset. ** Including two interviewers. Abbreviations: AWS - adults who stutter; CWS - children who stutter; PWS - people who stutter.}
\label{tab:scale_comparison}
\begin{tabular}{p{0.15\linewidth}|p{0.08\linewidth}|p{0.13\linewidth}|p{0.13\linewidth}|p{0.3\linewidth}|p{0.08\linewidth}}
\toprule
\textbf{Dataset}   & \textbf{Duration} & \textbf{Speakers}  & \textbf{Transcription} & \textbf{Tasks} & \textbf{Language} \\ \midrule
LibriStutter~\cite{libristutter} & 20 hrs & 50 non-PWS &  Yes & audiobook & English \\
UCLASS*~\cite{uclass} & 53 mins & 25 CWS & Yes & conversation & English \\
SEP-28k~\cite{sep-28k} & 23 hrs & not reported & No & podcast & English \\ 
FluencyBank*~\cite{fluencybank} & 3.5 hrs & 32 AWS &  Yes & conversation and reading & English\\
KSoF~\cite{bayerl_KSoFKasselState_2022} & 4.6 hrs & 37 PWS & No & spontaneous and reading & German \\ 
\textbf{AS-70} & \textbf{48.8 hrs} & \textbf{72 AWS**} & \textbf{Yes (verbatim)} & \textbf{conversation, voice commands} & \textbf{Mandarin} \\ 
\bottomrule
\end{tabular}
\vspace{-10pt}
\end{table*}

\begin{table*}[ht]
    \centering
    \caption{Distribution of annotated stuttering events. FluencyBank's annotation was done in another work~\cite{sep-28k}. \textbf{AS-70 Conversation} includes two interviewers.}
    \label{tab:stuttering_type_dist}
\begin{tabular}{l|c|c|ccccc} \toprule 
  & \textbf{Avg. Stuttering Rate}& \textbf{Total} & \multicolumn{5}{c}{\textbf{Event Type Distribution \%}} \\ 
 & \textbf{(per minute)} & \textbf{Stuttering Events} & \textbf{{[}{]}} & \textbf{/b} & \textbf{/p} & \textbf{/r} & \textbf{/i} \\ \hline
LibriStutter~\cite{libristutter} & 12.5 & 15,000 & 20.0 & 20.0 & 20.0 & 20.0 & 20.0 \\
SEP-28k~\cite{sep-28k} & 12.26 & 17,267 & 16.0 & 19.5 & 16.3 & 13.6 & 34.6 \\
FluencyBank\cite{fluencybank} & 13.88 & 2,875 & 15.0 & 14.9 & 11.8 & 19.1 & 39.3 \\
KSoF~\cite{bayerl_KSoFKasselState_2022} & 13.05 & 3,602 & 6.0 & 32.2 & 18.7 & 22.9 & 20.2 \\
\textbf{AS-70 Conversation}& 15.58 & 29,017 & 42.7 & 6.9 & 19.8 & 8.7 & 21.9 \\
\textbf{AS-70 Command} & 8.11 & 8,636 & 53.0 & 8.4 & 16.9 & 15.8 & 5.9 \\ \bottomrule
\end{tabular}
\vspace{-10pt}
\end{table*}

\begin{table*}[ht!]
\centering
\caption{Stuttering event distribution of SED task datasets. \textbf{no-dis}: no-disfluency. \textbf{AS-70 Conversation} includes two interviewers.}
\label{tab:sed_dist}
\begin{tabular}{l|c|c|cccccc} \toprule 
  & \textbf{Avg. length (s)}& \textbf{Number} & \multicolumn{6}{c}{\textbf{Event Type Distribution \%}} \\ 
 & & \textbf{of clips} & \textbf{{[}{]}} & \textbf{/b} & \textbf{/p} & \textbf{/r} & \textbf{/i} & \textbf{no-dis} \\ \hline
SEP-28k~\cite{sep-28k} & 3.00 & 28,177 & 9.8 & 12.0 & 10.0 & 8.3 & 21.2 & 56.9 \\
FluencyBank~\cite{fluencybank} & 3.00 & 4,144 & 10.38 & 10.33 & 8.16 & 13.25 & 27.27 & 54.10 \\
KSoF~\cite{bayerl_KSoFKasselState_2022} & 3.00 & 5,597 & 3.88 & 20.74 & 12.02 & 14.76 & 12.97 & 24.75 \\
\textbf{AS-70 Conversation}& 5.69 & 19,654 & 38.67 & 8.45 & 22.12 & 10.77 & 24.80 & 36.31 \\
\textbf{AS-70 Command} & 2.87 & 22,299 & 14.09 & 2.87 & 5.31 & 5.08 & 1.76 & 79.61 \\ \bottomrule
\end{tabular}
\vspace{-15pt}
\end{table*}

Table~\ref{tab:scale_comparison} presents a comparison of dataset scales between AS-70 and 4 other open datasets. A total duration of 48.8-hours speech data were included in AS-70 dataset from 70 recording sessions. Excluding two interviewers, each participant contributed on average 457 utterances and 33.0 minutes of total speech, with an average of 141 utterances and 17.8 minutes for conversations and an average of 316 utterances and 15.23 minutes for voice command reading.

Table~\ref{tab:scale_comparison} indicates that the AS-70 dataset is the largest among open datasets in terms of duration and participants. When compared to the previously largest open dataset, Sep-28k~\cite{sep-28k}, which is annotated at a 3-second clip level without transcription, the AS-70 dataset is twice as large, annotated at the character level, and includes verbatim transcription. Moreover, the AS-70 dataset encompasses both spontaneous conversational and voice command reading speech tasks, and stands out as the only non-Western language open dataset.

Table~\ref{tab:stuttering_type_dist} depicts the distribution of the stuttering events across four datasets. The \textit{Average Stuttering Rate} is calculated by dividing the total count of stuttering events by the duration of the speech. The \textit{Event Type Distribution} is calculated by the formula ``number of labels of a certain stuttering event type / overall number of all stuttering event labels". It's important to note that a direct comparison between Sep-28k and AS-70 may not be entirely equitable. This is because AS-70 dataset is annotated on the character level, offering finer granularity compared to the clip level annotation of Sep-28k. This difference may partially account for the higher average stuttering rate observed in the AS-70 conversation part compared to Sep-28k. 

Additionally, there is a significant shift towards more word and phrases repetitions and fewer sound repetitions in the AS-70 dataset, signaling potential phonological differences between stuttering in Chinese and in English. Meanwhile we notice that SEP-28k dataset contains almost twice more interjections than in AS-70 conversations, which could be attributed to different definitions of interjections in these two datasets. While SEP-28k includes any filler words, such as ``um,'' ``uh,'' and ``you know'' as stuttering interjections, our annotation excludes natural interjections that blend into the speech flow.

In line with previous studies~\cite{sep-28k,bayerl_KSoFKasselState_2022}, we split the utterances of the AS-70 dataset into short clips for the SED task. Table~\ref{tab:sed_dist} provides a comparison between the AS-70 dataset and other three datasets. Notably, AS-70 dataset contains 49\% more clips than the Sep-28k dataset.

To compute the \textit{Event Type Distribution}, we employ the consistent formula as in the work~\cite{sep-28k}: ``number of clips containing a certain stuttering event type / total number of clips". An interesting observation is that the conversation part of the AS-70 exhibits almost 4 times word/phrase repetitions and 2.2 times prolongation compared to Sep-28k. As a relatively easy task for our participants, the command reading task is reflected by a high no-disfluency rate, low interjection, and block rate.

\section{Experiments}\label{sec:experiments}

The objective of the work in this section is not to attain state-of-the-art results but to highlight the limitations of specific pre-trained models through an evaluation of the AS-70 dataset. Additionally, we aim to demonstrate the enhancements achievable by integrating our dataset into the model training process.

\vspace{-5pt}
\subsection{Data partition}
\begin{CJK*}{UTF8}{gbsn}
As highlighted in the study of Bayerl et al.~\cite{bayerl_InfluenceDatasetPartitioning_2022}, the data partition method can heavily impact the reliability of experimental results. A data partition that neglects speaker exclusivity can cause a model to learn speaker-specific traits, resulting in overly optimistic results. We ensure that speakers with varying degrees of stuttering severity are distributed across each partition.

The stuttering severity of a speaker is measured using the stuttering rate ($SR$), calculated by dividing the number of stuttering events by the number of non-stuttering characters in the transcription. E.g., the annotation ``嗯/i/p，我[我我]的名/b字是小/r明。" would lead to a stuttering rate of 71.42\% as there are 5 stuttering events and 7 non-stuttering characters. The stuttering severity is classified as mild: $SR <= 7\%$ for 45 participants, moderate: $7\% < SR <= 12\%$ for 16 participants, and severe: $ SR > 12\%$ for 9 participants. The speaker numbers of train/development/test partitions can be found in the link\footref{fn:webpage}. The data of two PWS interviewers is included in the train partition.
\end{CJK*}

\vspace{-5pt}
\subsection{ASR}\label{sec:asr}

\begin{table*}[ht!]
\caption{The CER(\%) results of different ASR model.}
\label{tab:results_asr}
\centering
\begin{tabular}{@{}lllllllll@{}}
\toprule
Model                         & Pre-trained Dataset     & Fine-tuned Dataset      & mild  & moderate & severe & conversation & command & all   \\ \midrule
                            & WenetSpeech            & -         & 11.41 & 17.81    & 33.21 & 11.82        & 20.40   & 15.00 \\
\multirow{-3}{*}{Conformer}   & WenetSpeech            & AS-70            & 5.20  & \textbf{8.00}    & 9.25 & \textbf{7.96}         & 3.32   & \textbf{6.24}  \\ \midrule
                                   & WenetSpeech (unlabel)            & AISHELL-1  & 16.28     & 22.86   & 26.76   & 16.80   & 23.79   & 19.06    \\
\multirow{-2}{*}{Hubert-large}     & WenetSpeech (unlabel)            & AS-70 & 6.25  & 9.67     & \textbf{7.85}  & 9.75         & \textbf{2.03}    & 7.25  \\ \midrule
                                   & Whisper                & -      & 14.50   & 28.50   & 95.33   & 17.83   & 46.85 & 27.20  \\
\multirow{-2}{*}{Whisper-large-v2} & Whisper                & AS-70    & \textbf{5.18}  & 13.46  & 18.92  & 10.19  & 5.74   & 8.75  \\ \bottomrule
\end{tabular}
\vspace{-15pt}
\end{table*}

We conduct an in-depth evaluation of ASR performance using the AS-70 dataset, focusing on three distinct models: Conformer~\cite{conformer}, HuBERT~\cite{hubert}, and Whisper~\cite{whisper}. These models are selected to represent a comprehensive range of approaches to ASR, including supervised end-to-end ASR, self-supervised pretraining, and large-scale semi-supervised methods. To emphasize semantic transcription, the annotations undergo preprocessing to exclude stuttering event labels, stuttering characters, and punctuation.


\textbf{Conformer.}
This system is based on Wenet~\cite{wenet}. U2++ is a unified two-pass framework with bidirectional attention decoders, which includes the future contextual information by a right-to-left attention decoder to improve the representative ability of the shared encoder and the performance during the rescoring stage.
We show the results of the u2++ conformer model pre-trained using WenetSpeech dataset~\cite{zhang2022wenetspeech} and the results of the fine-tuned model with AS-70 train partition.

\textbf{Hubert.}
HuBERT has demonstrated remarkable efficacy in ASR through self-supervised learning. We employ the Chinese Hubert Large model as our pre-trained experiment system~\cite{chinesehubert}. We use the results of AISHELL-1~\cite{aishell_2017} fine-tuning as a baseline to compare with the model fine-tuned by the AS-70 dataset.

\textbf{Whisper.}
Whisper exhibits exceptional proficiency across multiple languages, developed through semi-supervised training on a large-scale dataset. Our baseline employs the Whisper large-v2 model for direct inference. Subsequent efforts are channeled towards fine-tuning the Whisper model, specifically with the AS-70 dataset.

\textbf{Results and discussion.}
Table \ref{tab:results_asr} indicates the above models' Character Error Rate (CER) results against the AS-70 test set. These results communicate a clear message: even though the above models have performed impressively with generic data, their proficiency in dealing with stuttering data seemed to falter. However, after fine-tuning using AS-70 data, all models exhibit commendable improvement. The conformer with AS-70 fine-tuning takes the lead, possibly owing to extensive labeled data in the conformer pretrain stage. On the other hand, Whisper seems to falter on severe data. Upon examination, we notice that Whisper tends to generate an overabundance of repeat labels when decoding utterances from severe-level PWS.

An interesting finding from these experiments is the relatively low CER registered on command data processed by the AS-70 fine-tuned models, possibly due to the training set containing identical command text as that in the test set, with only variations being the speakers. We plan to address this limitation in the future work by considering the factor of command text in the data partition process. Examples of ASR results can be found in the link\footref{fn:webpage}.

\vspace{-5pt}
\subsection{SED}
The evaluation in SED is undertaken by establishing a fundamental benchmark predicated on random guessing, thus providing a comparative baseline for subsequent analyses. We employ some reputed SED methods, such as StutterNet~\cite{stutternet}, ConvLSTM~\cite{sep-28k}, Conformer and wav2vec2.0~\cite{Bayerl2022DetectingDI}, to assess the effectiveness of these methods.

\textbf{StutterNet.} 
StutterNet~\cite{stutternet} uses a time-delay neural network suitable for capturing contextual aspects of the disfluent utterances, which is trained on the MFCC input features. We reproduce the structure in Stutternet~\cite{stutternet} with 12.2M parameters. We use multi-task learning with two output branches: a fluent/dysfluent prediction and a soft prediction for each of the five event types.

\textbf{ConvLSTM.} 
ConvLSTM~\cite{sep-28k} input is a set of 40-dimensional mel-filterbank energy features. Feature maps from the convolution layer are combined after batch normalization and fed to three LSTM layers, which result in a 1.6M parameter size. We use the same multi-task learning as in StutterNet.

\textbf{Conformer.}
The same Wenet conformer~\cite{conformer} encoder architecture mentioned in the section~\ref{sec:asr} is applied. However, due to the small amount of training data, we use 3 conformer blocks, which result in 9.7M parameter size. We use single-task learning, which is to predict the five event types. The multi-label soft margin loss is used for the model training. 

\textbf{Wav2Vec2.0.} 
Referring to Bayerl el al.'s approach~\cite{Bayerl2022DetectingDI}, we fine-tune the wav2vec2.0 base model. The model we used for our initial experiments (Chinese-wav2vec2-base)~\cite{chinesew2v2} is pretrained in an unsupervised manner on the WenetSpeech corpus. We use the same single-task learning as in Conformer.
\vspace{-5pt}

\begin{table}[ht!]
\caption{The F1-score(\%) of different SED model. We report results for each label individually.} 
\label{tab:results_sed}
\centering
\begin{tabular}{@{}llllll@{}}
\toprule
Model      & /p    & /b    & /r    & {[}{]} & /i   \\ \midrule
Random guess     & 15.65 & 8.07  & 15.34 & 31.7    & 17.03  \\
StutterNet~\cite{stutternet} & 61.07 & 33.33 & 47.81 & 50.12   & 58.82  \\
ConvLSTM~\cite{sep-28k}   & 33.30 & 18.22 & 30.19 & 64.02   & 46.70   \\
Conformer~\cite{conformer}  & 66.77 & 30.94 & 46.84 & 65.49   & 73.10  \\
Wav2vec2.0~\cite{Bayerl2022DetectingDI} & \textbf{70.48}  & \textbf{42.51} & \textbf{65.76} & \textbf{78.48}   & \textbf{83.80} \\ \bottomrule
\end{tabular}
\end{table}
\vspace{-10pt}

\textbf{Results.}
Table \ref{tab:results_sed} indicates the F1-score of the above models for the five stuttering event types. All models exceed the random guess baseline in each event type. ConvLSTM underperformed in comparison, which is attributed to its limited parameter count of 1.6m. StutterNet and Conformer display intermediate results, whereas wav2vec 2.0 emerged as the most proficient, benefiting significantly from its pre-training on a large dataset. It is worth noting that the block (/b) performance is substantially worse than the other stuttering types, probably attributed to the fact that blocks are often represented solely by silence in the speech. Similar findings are reported in Lea et al.~\cite{Lea_chi_2023} and Bayerl et al.~\cite{bayerl_InfluenceDatasetPartitioning_2022}. For a detailed SED results on stuttering severities and speech tasks, please refer to the link\footref{fn:webpage}.

\vspace{-5pt}
\section{Conclusion}
This paper introduces the AS-70 dataset, which is the first publicly available Mandarin stuttered speech dataset and is notable for being the largest in its category. AS-70 consists of conversational and voice command reading speech recordings with verbatim manual transcription, making it suitable for various speech-related tasks. In addition, baseline systems have been established, and experimental results are presented for ASR and SED tasks, demonstrating significant improvements in state-of-the-art ASR models by incorporating this dataset into model fine-tuning processes. We hope that AS-70 can assist in detecting stuttering and aiding people who stutter in developing speech interaction systems.


\clearpage

\bibliographystyle{IEEEtran}
\bibliography{template}

\end{document}